\documentclass[aps,prl,twocolumn,superscriptaddress,showpacs,english]{revtex4-1}

\usepackage[T1]{fontenc}
\usepackage{lipsum}

\usepackage[latin9]{inputenc}
\usepackage{babel}
\usepackage{amsmath}
\usepackage{amssymb}
\usepackage{wasysym}
\usepackage{graphicx}
\usepackage{multirow}
\usepackage{gensymb}
\usepackage[dvipsnames]{xcolor}
\usepackage{pifont}
\usepackage{microtype}
\usepackage{xcolor}
\usepackage{soul}
\usepackage[version=3]{mhchem}
\usepackage{chemformula}[2014/04/08] 
\usepackage{booktabs}    
\usepackage{dcolumn}     
\usepackage{bm}          
\usepackage{endnotes}

\usepackage[linktocpage=true,
  colorlinks=true, 
  pdfborder={0 0 0},
  linkcolor=blue,
  citecolor=red,
  filecolor=yellow,
  urlcolor=blue,
  bookmarks,
  pdfauthor={},
]{hyperref}

\usepackage{longtable}

\newcommand{\Unidonostia}{Fisika Aplikatua 1 Saila, Gipuzkoako Ingeniaritza Eskola, University of the Basque Country (UPV/EHU), Europa Plaza 1, 20018 Donostia/San Sebasti\'an, Spain}
\newcommand{\CFM}{Centro de F\'isica de Materiales (CSIC-UPV/EHU), Manuel de Lardizabal Pasealekua 5, 20018 Donostia/San Sebasti\'an, Spain}
\newcommand{\DIPC}{Donostia International Physics Center (DIPC),  Manuel de Lardizabal Pasealekua 4, 20018 Donostia/San Sebasti\'an, Spain}
\newcommand{\Sorbonne}{Laboratoire de Chimie Th\'eorique (LCT), Sorbonne Universit\'e CNRS, 75005 Paris (France)}

\begin{document}

\title{Strong correlation between bonding network and critical temperature in hydrogen-based superconductors}

 
\author{Francesco Belli}
\affiliation{\CFM} \affiliation{\Unidonostia}

\author{J. Contreras-Garcia}
\affiliation{\Sorbonne}

\author{Ion Errea}
\affiliation{\CFM} \affiliation{\Unidonostia} \affiliation{\DIPC}

\begin{abstract}
Recent experimental discoveries show that hydrogen-rich compounds can reach room temperature superconductivity, at least at high pressures. Also that there exist metallic hydrogen-abundant systems with critical temperatures of few Kelvin, or even with no trace of superconductivity at all. By analyzing through first-principles calculations the structural and electronic properties of more than one hundred compounds predicted to be superconductors in the literature, we determine that the capacity of creating a bonding network of connected localized units is the key to enhance the critical temperature in hydrogen-based superconductors, explaining the large variety of critical temperatures of superconducting hydrogen-rich materials. We define a magnitude named as the {\it networking value}, which correlates well with the predicted critical temperature, much better than any other descriptor analyzed thus far. This magnitude can be easily calculated for any compound by analyzing isosurfaces of the electron localization function. By classifying the studied compounds according to their bonding nature, we observe that the {\it networking value} correlates with the critical temperature for all bonding types. Our analysis also highlights that systems with weakened covalent bonds are the most promising candidates for reaching high critical temperatures. The discovery of the positive correlation between superconductivity and the bonding network offers the possibility of screening easily hydrogen-based compounds and, at the same time, sets clear paths for chemically engineering better superconductors.
\end{abstract}

\maketitle


\section{Introduction}

The field of hydrogen-based superconductivity has progressed enormously since 1968, when Ashcroft first proposed that pressurized hydrogen may become a high temperature superconductor \cite{1}. While the first discovered superconductors were Th$_4$H$_{15}$ \cite{2} in 1970 and PdH \cite{2B} in 1972, with not very promising critical temperatures ($T_c$) of 7.6 K and 5 K respectively, the more recent experimental discoveries at high pressures show that superconductivity on hydrogen-based compounds can span from a few Kelvin to room temperature. Few examples are the recently synthesized PrH$_9$\cite{4F}, where $T_c=7$ K at 125 GPa, AlH$_3$ where no superconductivity was observed above 4 K despite the predictions \cite{4E}, and, on the other extreme, H$_3$S \cite{DrozdovEremets_Nature2015}, YH$_9$ \cite{4D,snider2020superconductivity}, YH$_6$ \cite{troyan2020anomalous}, and LaH$_{10}$ \cite{4,4B} all reaching critical temperatures well above 200 K at megabar pressures. In addition, the recent observation of a $T_c$ of 288 K at 267 GPa in a compound formed by sulfur, carbon, and hydrogen \cite{4C} confirms that hydrogen-based superconductors can be room-temperature superconductors. This observation finally dismisses the maximal limits for $T_c$ initially suggested by Cohen and Anderson for electron-phonon driven superconductivity \cite{9}. Hydrogen-based compounds are thus the best currently available candidates to reach ambient temperature and pressure superconductivity. 

In order to discover new compounds with high $T_c$ at low pressures, a simple physical-chemical understanding of the properties enhancing the critical temperatures in hydrogen-based systems is necessary. In this regard, the hundreds of compounds predicted to be superconductors with first principles crystal structure prediction techniques constitute a rich working dataset to extract conclusions \cite{7,15,doi:10.1146/annurev-conmatphys-031218-013413}. Among these predictions, the highest $T_c$ values are 300 K for pure metallic hydrogen \cite{5,6} and 326 K for the YH$_{10}$ binary compound \cite{8}. Attempts have been made to increase $T_c$ further through ternary compounds, for instance with H$_3$S$_{1-x}$P$_x$ \cite{10} and Li$_2$MgH$_{16}$ \cite{11}. Aiming at extracting useful information from this dataset, two main routes are being explored: on the one hand, machine learning methods \cite{12,13,14} are starting to be employed to further increase the list of predicted systems, although the obtained new compounds so far do not beat the already known; on the other hand, additional efforts are being invested into classifying these superconductors using simple footprints based on structural, chemical, and electronic properties \cite{13,15,17}. These studies suggest hydrogen rich systems with highly symmetrical structures and high density of states (DOS) at the Fermi level are the best candidates for high-temperature superconductivity.

Even if these properties are able to suggest good trends, they serve necessary but not sufficient conditions. This ultimately means there are not good optimizers thus far: improving these parameters will not necessarily lead to an improvement of the superconducting critical temperatures. In other words, even if the footprints for a good superconductor are somewhat clear, we cannot yet rely on simple variables to estimate the superconducting temperatures, clarify the reason for such a broad spectrum of $T_c$ values, and, ultimately, chemically engineer better superconductors.

In this work we investigate the chemical, structural, and electronic properties through {\it ab initio} methods based on density functional theory (DFT) for a set of 178 hydrogen-based superconductors previously predicted in the literature \cite{7}, including pure hydrogen and binary compounds. Our ultimate goal is to provide a simple understanding of the origin of the high $T_c$ in these compounds. We focus mainly on the electronic and structural properties by means of chemical bonding descriptors, hydrogen-hydrogen distance, electronic charge, and density of states at the Fermi level. We review the impact on the predicted $T_c$ of many of these descriptors, which have been already somewhat studied on a case to case basis in the literature but do not reveal conclusive. We identify that the electron pairing and delocalization play instead the final role. This takes us to propose a universal descriptor based on the identification of electronic delocalization networks, identified by means of the electron localization function (ELF) \cite{18,19,20,21,22}, which quantifies the network of the electronic bonding. We define a simple magnitude, the {\it networking value}, which is easily obtained from the calculation of ELF isosurfaces. Such quantity reveals useful to have a first estimate of the superconducting critical temperature without the need of performing electron-phonon coupling calculations. To the best of our knowledge, it is the first time that such a descriptor is proposed in the literature. We believe the analysis of the {\it networking value} is the first step for a high-throughput analysis of potential new high-$T_c$ systems. It also provides clear paths for chemically engineering better hydrogen-based superconductors, guiding the quest for high-$T_c$ compounds among the vast possibilities offered by ternary compounds. 

\section{Results}

\noindent
\textbf{Chemical composition and bonding categories.}
In order to provide a comprehensive understanding of the different type of hydrogen-based compounds predicted to exist in the literature, it is convenient to categorize them in families according to the nature of the chemical bonding of the hydrogen atoms in the system. Our classification is guided by the study of the ELF \cite{18,19,20,21,22} and the atomic charge distribution obtained through the Bader analysis \cite{23,24,25,26} (see Methods for more details). The ultimate goal of this investigation is to better understand what kind of chemical interaction is the most beneficial for superconductivity among these compounds.

\begin{figure*}[ht]
\includegraphics[width=\textwidth]{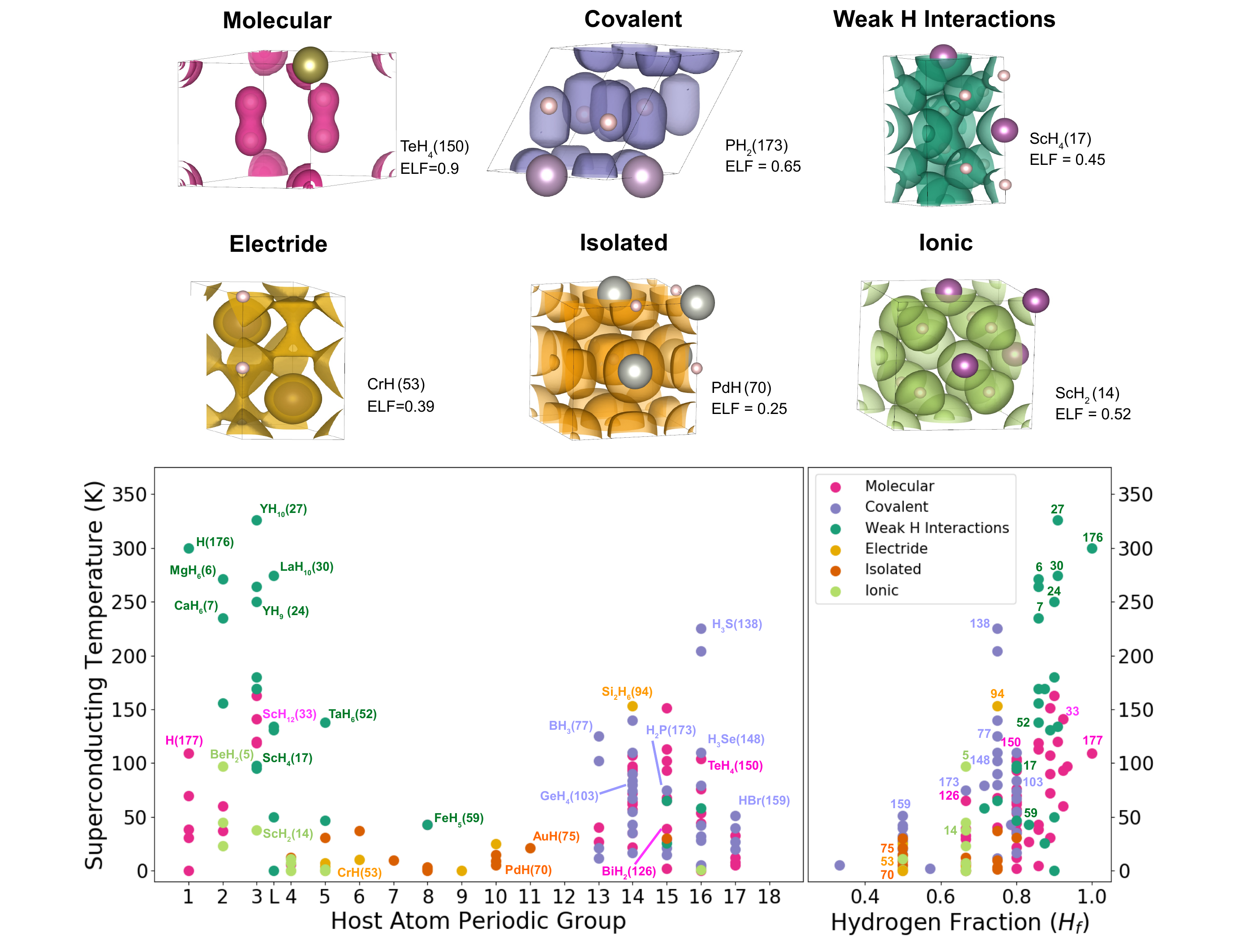}
\caption{\textbf{Classification of hydrogen-based superconductors according to their bonding categories.} The top panel shows representative systems for the different categories with ELF isosurfaces at different values: TeH$_4$ for {\it molecular} systems (magenta),  PH$_2$ for {\it covalent} systems (purple), ScH$_4$ for systems with \textit{weak covalent hydrogen-hydrogen interactions} (dark green), CrH for {\it elecrtides} (yellow), PdH for {\it isolated} systems (orange), and ScH$_2$ for {\it ionic} systems (light green). The lower panel shows respectively the critical temperature as a function of the host atom periodic group (left panel) and the hydrogen fraction in the compounds (right panel). In the left panel, the name of some compounds is explicitly given together with the index (in parenthesis) given to each compound in Extended Data Tables I - III. In the right panel, and also in Figs. \ref{fig:Electronic} and \ref{fig:Network}, just the index is given for these compounds for the sake of brevity.}
\label{fig:Group}
\end{figure*}

After a thorough analysis of the ELF in our test set, we identify six different families according to the nature of the chemical bonding, namely, \textit{molecular} systems, \textit{covalent} systems, systems driven by \textit{weak covalent hydrogen-hydrogen interactions}, systems with \textit{electride} behavior, \textit{ionic} systems, and \textit{isolated} systems. In each case, the nature of the bonds is identified through the analysis of the ELF saddle points between different atoms. Since bonding properties are mainly local, each system can belong to more than just one family. However, in order to simplify the analysis, we focus on the most dominant feature for each compound. A representative for each family, together with the distribution of the families through the groups of the periodic table and the amount of hydrogen fraction in each case, is shown in Fig. \ref{fig:Group}. The hydrogen fraction ($H_f$) takes the following form:
\begin{equation}
    H_f = \frac{N_H}{N_H +N_X},
\end{equation}
where $N_H$ and $N_X$ are the number of hydrogen and non-hydrogen atoms in the primitive cell, respectively.

The \textit{molecular} family describes all systems having at least one pair of hydrogen atoms forming a molecule. The latter can be identified through the ELF analysis by locating an isosurface surrounding an isolated molecule at very high values of ELF (see the magenta surface around the hydrogen molecules in TeH$_4$ in Fig. \ref{fig:Group}). A system is chosen to be molecular if an isolated pair of hydrogen atoms appears connected at a value of ELF higher than 0.85, i.e., if the minimum ELF value in between the hydrogen atoms is above 0.85. Systems with {\it molecular} behavior appear between groups 1 to 4 and 13 to 17. All of them have a value of $H_f>0.\overline{6}$, meaning that in the unit cells there are at least two hydrogen atoms per host. We note that {\it  molecular} systems tend to have very high values of $H_f$, which reflects that in most cases several molecules exist per host atom. The highest critical temperatures for these {\it  molecular} compounds have been predicted for ScH$_{12}$(33) and TeH$_4$(150), with $T_c$ values around 150 K (the number in parenthesis after a given compound corresponds to the index given for each compound in Supplementary Data Tables I - III). The critical temperatures of the {\it molecular} compounds span from few Kelvin to these high values. 

The \textit{covalent} family is composed of systems where covalent bonds between hydrogen and the host atoms are dominant. Throughout our discussion we shall label the host atom as X, with X$\neq$H. For H-X bonds, the covalent character can be identified by an elongation toward the host atom of the ELF isosurface surrounding the hydrogen (i.e. a polarized covalent bond). This is exemplified by the purple surfaces pointing from the H (small spheres) to the P atoms (big spheres) in PH$_2$ in Fig. \ref{fig:Group}. {\it Covalent} systems appear for groups 13 to 17, and are mostly related to the host atom's $p$ type orbital character. Some of the highest critical temperature for these systems have been predicted for H$_3$S(138) at 200 K, BH$_3$(77) at 125 K, and H$_3$Se(148) at 110 K.

The \textit{weak covalent hydrogen-hydrogen interaction} family is dominated by compounds with predominant weak H-H covalent interactions. The difference with the molecular case is that hydrogen molecules or clusters appear elongated or quasi dissociated. This is illustrated with the dark green surfaces in ScH$_4$ in Fig. \ref{fig:Group}, where bonds between hydrogen atoms appear at much lower values of ELF. From a quantitative point of view we assume that a group of hydrogen atoms is weakly bonded if the ELF at the bond point is within the range [0.4 - 0.85]. These systems mostly appear between groups 1 to 5. In this bonding type, interactions seem to be purely related to hydrogen atoms whilst the host atoms appear as inert or acting as a chemical pre-compressor or electron donor. Compounds with this bonding characteristic tend to contain many hydrogen atoms per host, as it happens for the molecular compounds. For compounds with the host atom in a low group of the periodic table, the host atoms valence electrons are donated to the hydrogen atoms resulting in a weakening of the H-H bonds, which translates into an increment of the H-H distance. This family shows the highest predicted critical temperatures, the highest being 326 K for YH$_{10}$(27) and 300 K for metallic hydrogen H(76).

The \textit{ionic} family is formed by those systems whose hydrogen atoms show an ionic character. This is identified by an isolated proto-spherical ELF isosurfaces surrounding the hydrogen atoms. In addition, for a system to be hereby considered as \textit{ionic} the mean extra charge per hydrogen atom must be more than 0.5 electrons. This is illustrated by the spherical light green surfaces around hydrogen in ScH$_2$ in Fig. \ref{fig:Group}, where the charge on H is 0.67 electrons. Ionic behavior was observed between groups 2 to 5 of the periodic table, in all cases with low values of $H_f$. The ionic bonding originates from the strong difference in electronegativity between host and hydrogen atoms, which is always increased under pressure \cite{31}. This family transforms into the \textit{weak covalent hydrogen-hydrogen interaction} class when $H_f$ increases, since the donated charge becomes small in respect to the number of recipient hydrogen atoms. Critical temperatures for these systems are low, with the highest being 45 K for BeH$_2$(4).

The \textit{electride} family contains systems featuring electride behavior, i.e. compounds with electrons localized in the voids. The latter can be identified by isolated pockets of localized electrons in empty space of the crystal as the ones shown in yellow for CrH in Fig. \ref{fig:Group}. From a quantitative viewpoint, electride behavior is characterized in terms of isolated isosurfaces not surrounding any nuclei with ELF maximum values in between 0.35 to 0.7. Note that metallic compounds are also included in this family. Metallic cases also show isolated bubbles of ELF occupying the voids, but their profile is flatter. Given the difficulties to set a quantitative barrier, we have merged them in a unique family. Electrides and metals only appear mainly between groups 5 and 10, and are among the systems with the lowest value of $H_f$ reaching a maximum of 3 hydrogen atoms per host ($H_f = 0.75$). Critical temperatures for these systems are low, not reaching above 50K, exception made for the Si$_2$H$_6$(94) with a $T_c$ of 153 K, where the electride-metallic behavior seems to be arising from H-Si covalent bonds (not justifying a further differentiation between electrides and metals).

The remaining family includes all the materials featuring extremely weak bonds between hydrogen and host atoms. These systems have been named as \textit{isolated} and are identified by the lack of any kind of connection of the ELF isosurfaces above 0.25. These systems have low critical temperatures not reaching above  40 K and appear mainly between groups 5 and 12 of the periodic table. They also show a weak capacity of hosting a large number of H atoms per X atom, as they have the lowest values of $H_f$. 

Overall our results highlight that characterizing the bonding type of a solid thanks to these families enables to discard a great number of compounds as potential high-$T_c$ compounds. Covalent interactions, be it weak H-H or X-H are the most favorable for high-temperature superconductivity. This allows to identify the potential interesting combination of elements, especially with respect to the increasing search among ternary compounds.  The lowest $T_c$ values appear for {\it electrides} and {\it isolated} compounds, mostly appearing between groups 5 and 12, which do not show lots of potential as high-$T_c$ compounds. These families also show the lowest values of $H_f$.
\newline

\begin{figure*}[ht]
\includegraphics[width=\textwidth]{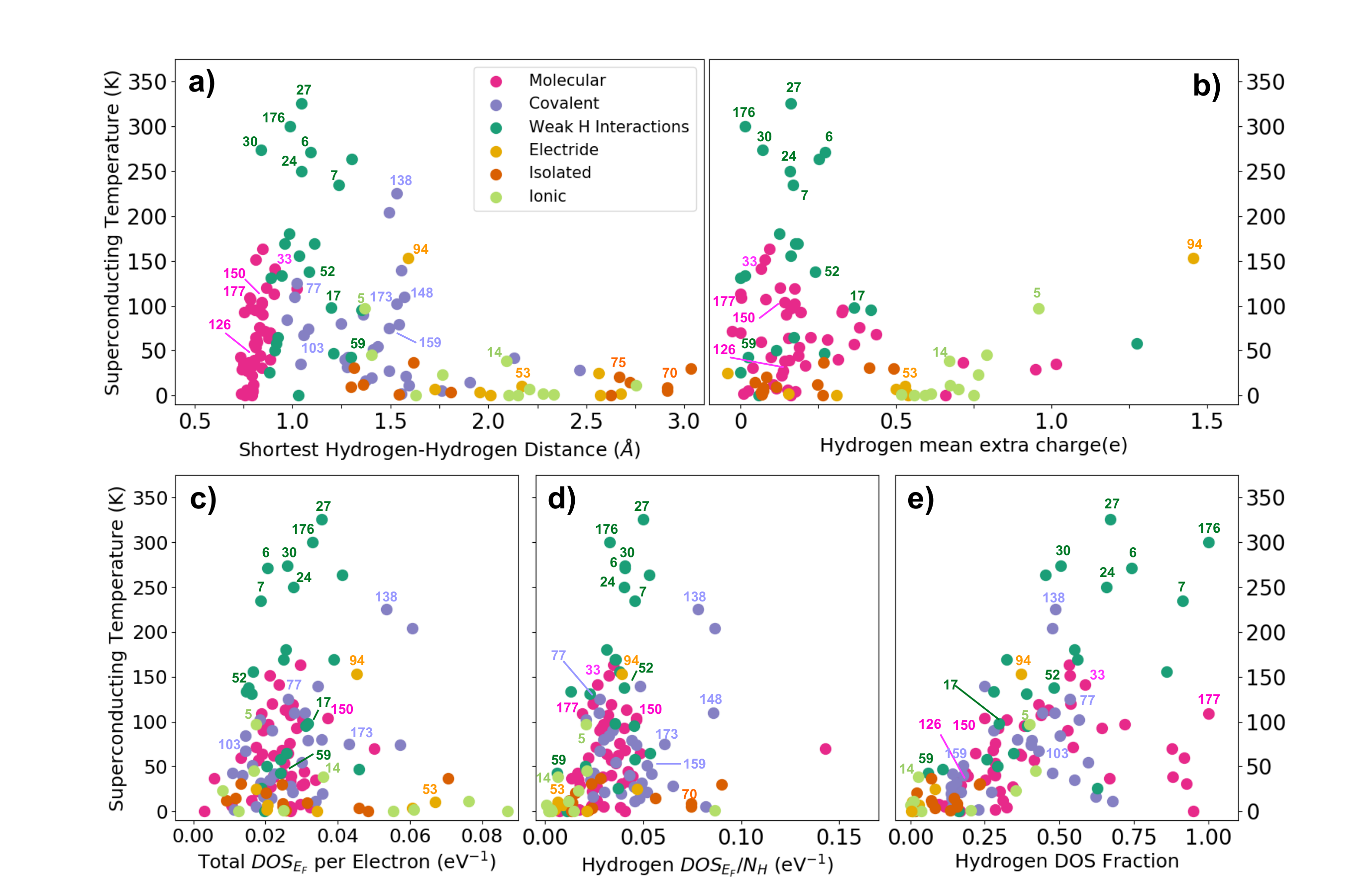}
\caption{\textbf{Structural and electronic properties.} Panel \textbf{a)} shows $T_c$ as a function of the shortest hydrogen-hydrogen distance for all the compounds. Panel \textbf{b)} shows $T_c$ as a function of the mean extra electrons per hydrogen atom. Panels \textbf{c)}, \textbf{d)}, and \textbf{e)}, respectively, show $T_c$ as a function of the total DOS at the Fermi level, the DOS at the Fermi level projected on the hydrogen $s$ orbitals per hydrogen, and the fraction of the total DOS at the Fermi level coming from the hydrogen orbitals.}
\label{fig:Electronic}
\end{figure*}
\noindent
\textbf{Hydrogen-hydrogen distance and electronic properties.}
After categorizing the different bonding families of hydrogen-based superconductors, we focus on understanding the trends of the predicted $T_c$ with structural and electronic properties. The results are summarized in Fig. \ref{fig:Electronic}. The analysis focuses on the shortest hydrogen-hydrogen distance for each compound, the charge distribution on hydrogen atoms, and the density of states at the Fermi level with the ultimate goal of finding correlations between such descriptors and the critical temperature.

Even if no general trend is observed when plotting $T_c$ as a function of the H-H shortest distance, several conclusions can be drawn. The structural analysis highlights an increment of the superconducting critical temperature with the increase of the shortest hydrogen-hydrogen distance for those systems where the bonding is driven by pure hydrogen interactions, meaning the {\it molecular} and \textit{weak covalent hydrogen-hydrogen interaction} families. For these two families the H-H distance spans from 0.74 \AA~for systems with $T_c$ below 1 K to a maximum of around 1.35 \AA~for the compounds with highest critical temperatures. In the region between 0.9 and 1.35 \AA~lie the currently predicted compounds with the highest superconducting temperatures reaching values as high as 300 K. In other words, elongated H-H interactions promote $T_c$. Interestingly, our analysis highlights that such H-H distance variation is not related to a variation of pressure, meaning that on a broad level, increasing the pressure does not necessarily yield an increment of the bonding distance, so composition rather than pressure would be a more relevant variable to tweak (see Supplementary Data Table I to see the pressure at which each compound is studied).

\begin{figure*}[ht]
\includegraphics[width=\textwidth]{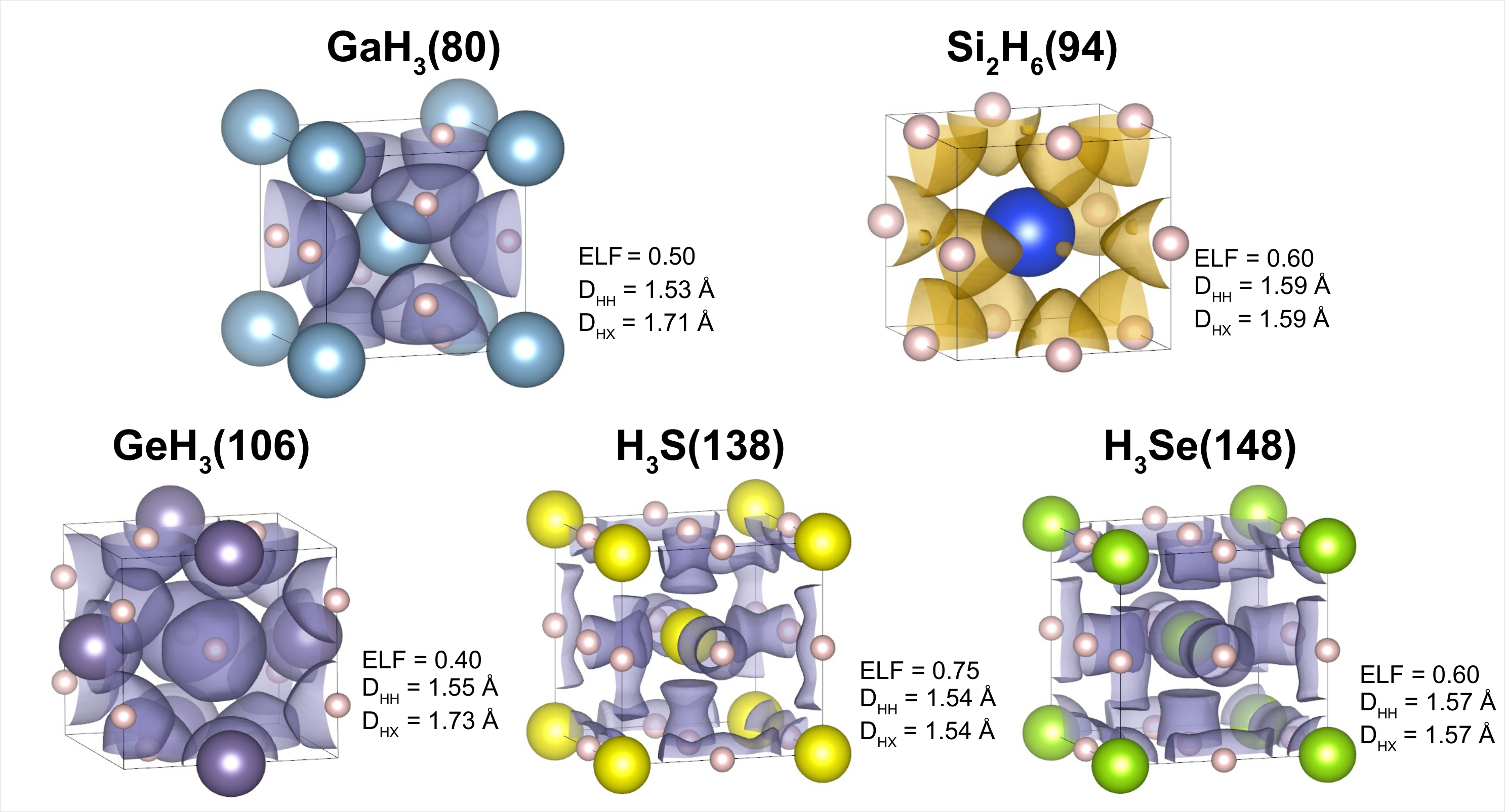}\\
\caption{\textbf{Highly symmetric high-$T_c$ systems .} This figure shows five compounds related to the $T_c$ spike around a hydrogen-hydrogen distance of 1.55 \AA . The pink spheres refer to hydrogen atoms, while the rest refer to the respective host atoms. An ELF isosurface is depicted in each case. Also the shortest H-H ($D_{HH}$) and H-X ($D_{HX}$) distances are noted. Most of these compounds belong to the {\it covalent} family and show a purple ELF isosurface. The Si$_2$H$_6$ compound is an {\it electride} (note the small ELF bubbles in empty sites) and its ELF isosurface is shown in orange.}
\label{fig:HHHComp}
\end{figure*}

The shortest hydrogen-hydrogen distance for the {\it covalent} family spans between 1 and 2.5 \AA. Low symmetry {\it covalent}  systems appear between 1 to 1.45 \AA\ where the short H-H distance is due to inter hydrogen bonds appearing beside the dominant hydrogen-host bonds. The highest reported superconducting temperature for these systems is about 135 K for BH$_3$(77). Interestingly, at around 1.55 \AA\ the {\it covalent} family shows a sharp spike in the predicted $T_c$ through systems sharing linear $X=H=X$ bonds originating through the host $p$ orbitals, a $\bar{3}$m point group, a value of $H_f$ equal to 0.75, and a lack of direct hydrogen-hydrogen bonds. Here lay systems as H$_3$S(138), H$_3$Se(148), GaH$_3$(80), and GeH$_3$(106). Interestingly, the Si$_2$H$_6$(94) {\it electride} also shares these features even if the bonding nature is slightly different. All these compounds are shown in Fig. \ref{fig:HHHComp} with a representative ELF isosurface. For H-H distances above approximately 1.55 \AA\ a sharp drop in $T_c$ appears, with systems not reaching above 50 K. This zone lacks direct hydrogen-hydrogen bonds and features systems with a low percentage of hydrogen, with mostly {\it isolated}, {\it electride}, and {\it ionic} behavior. Hence, covalent elongated bonds, H-H or H-X, with high $H_f$ seem to be the best candidates to increase $T_c$.

\noindent

Panel \textbf{b} of Fig. \ref{fig:Electronic} shows the mean extra electron per hydrogen atom obtained through the analysis of the Bader charge (see Methods). The {\it covalent} family has been excluded from this panel due to the unreliability of the Bader anlysis, which arises from the difficulties in assigning shared electrons in the H-X bonds to each atom. We estimate the mean extra electron per hydrogen atom ($\bar{\rho} $) as
\begin{equation}
   \bar{\rho} = \frac{\sum_{i=1}^{N_H} Q_i - N_H}{N_H},  
\end{equation}
where $Q_i$ is the number of electrons assigned to the $i$th hydrogen atom by the Bader analysis, while $N_H$ is the total number of hydrogen atoms in the primitive cell.

Overall hydrogen atoms tend to gain electrons due to their higher electronegativity with respect to other atoms in the cell. We observe that the highest $T_c$ values are associated with small charge transfers, i.e. from 0 to 0.25 extra electrons per hydrogen atom. As the extra charge increases, $T_c$ drops sharply below 50 K, exception made for the BeH$_2$(5), compound belonging to the {\it ionic} family with $T_c=97$ K, and  Si$_2$H$_6$(94), with $T_c=153$ K  and part of the {\it elecrtide} family.

It is often mentioned in the literature that an increment of (negative) charge on the hydrogen atom leads to a weakening of the hydrogen-hydrogen bonds \cite{11,104,112,https://doi.org/10.1002/wcms.1330}. Our results show that both {\it molecular} and \textit{weak covalent hydrogen-hydrogen interaction} families show non-negligible extra electrons on the hydrogen atoms. We also observe that the extra charge on the hydrogen is responsible for a slight increment of the H-H distance for the \textit{weak covalent hydrogen-hydrogen interaction} family. In fact, compounds within this family with a shorter H-H distance tend to have less extra electrons per hydrogen atom.  The maximum extra electron per hydrogen among this family is around 0.5, a value that increases up to 1 in the {\it molecular} systems.

Panels \textbf{c}, \textbf{d}, and \textbf{e} in Fig. \ref{fig:Electronic} report the results for the DOS analysis. A simplified equation for the electron-phonon coupling constant $(\lambda)$ may be given by \cite{PhysRev.167.331} 
\begin{equation}
 \lambda = \frac{N(E_F)\langle I^2 \rangle}{M\langle \omega^2 \rangle},
\end{equation}
where $N(E_F)$ is the DOS at the Fermi energy, $\langle I^2 \rangle$ is a term that is related to the electron-phonon matrix elements, $M$ is the mean mass of the primitive cell, and $\langle \omega^2\rangle$ is the mean quadratic phonon frequency. This suggests that a higher $N(E_F)$ should give rise to a higher electron-phonon coupling and, thus, also a higher superconducting critical temperature. Although this is a favorable condition for good superconductors, the results for the DOS prove this condition not to be univocally sufficient \cite{16}. The total DOS per electron at the Fermi level shown in panel \textbf{c} of Fig. \ref{fig:Electronic} shows a sharp increment in the highest superconducting temperatures for a DOS value of around 0.015 $eV^{-1}$. However, for such values of the DOS, compounds with very low $T_c$ can still be found, making the total DOS at the Fermi level not a good descriptor of high-$T_c$ compounds.

Considering that H atoms due to their light mass are responsible for the large values of $T_c$ in these compounds, we analyze the projection of the DOS at the Fermi level onto hydrogen atoms (see panel \textbf{d} of Fig. \ref{fig:Electronic}). However, the trends are similar to those obtained for the total DOS. In this case, the chance to have a higher superconducting critical temperature is related to an increment of DOS up to a value of 0.06 $eV^{-1}N_H^{-1}$. In particular, our results suggest that the key for high superconductivity is not strictly related to the value of the DOS. Instead, the fraction of active hydrogen atoms at the Fermi energy (see panel \textbf{e}) seems to be more relevant. In agreement with our previous findings, only {\it molecular} and \textit{weak covalent hydrogen-hydrogen interaction} families are able to reach high amounts of DOS coming from the hydrogen active atoms. This is directly related to the fact that the H-H bonds in those systems have a great contribution to the HOMO and LUMO. Nevertheless, false positives appear where the $T_c$ appears very low despite the large contribution of hydrogen states to the Fermi surface. In fact, the DOS misses the information on how  electrons are coupled with the lattice vibrations. Systems with very high values of the DOS at the Fermi energy but with very low electron-phonon coupling will not exhibit high superconducting critical temperatures.

\noindent
\textbf{Networking through the ELF.}
In order to screen new superconductors and guide the quest for new superconducting compounds, we need to look for an easily computable variable that will tell us not only about the interatomic bonding properties, but also how prone the system is to electron-phonon coupling. Even if some trends can be observed as discussed above, the descriptors analyzed so far in Figs. \ref{fig:Group} and \ref{fig:Electronic} are unable to capture when electrons couple more strongly to lattice vibrations and, thus, do not correlate with $T_c$. In this section we propose a new observable based on the study of electron (de)localization that, instead, correlates well with the predicted superconducting critical temperature.

The ELF is a proper function to analyze the degree of electronic localization as high value isosurfaces of the ELF reveal regions in space where electrons localize. In fact, for isosurfaces with values close to 1 the electrons are localized generally on atomic sites, and they start to delocalize toward neighbors and form bonds as the ELF value decreases (see Methods). In order to analyze this delocalization on a crystal size scale, we define the \textit{networking value} $\phi$ as the highest value of the ELF that creates an isosurface spanning through the whole crystal in all three cartesian directions. The $\phi$ value can thus be easily extracted by calculating the ELF and determining at which value a crystal sized isosurface is created when lowering the ELF value from 1. This isosurface  encloses most of the atoms in the crystal, but not necessarily all. The ELF saddle points reveal crucial for the determination of the {\it networking value}, especially for hydrogen based compounds. For these systems, where hydrogen-hydrogen bonds are dominant, the ELF saddle points identify the interatomic bonds that pave the crystal sized network. Thus, for the determination of the {\it networking value} it is sufficient to identify the ensemble of ELF saddle points at the highest value of ELF able to bridge the gap between different atoms and create the 3D network.

\begin{figure*}[ht]
\includegraphics[width=\textwidth]{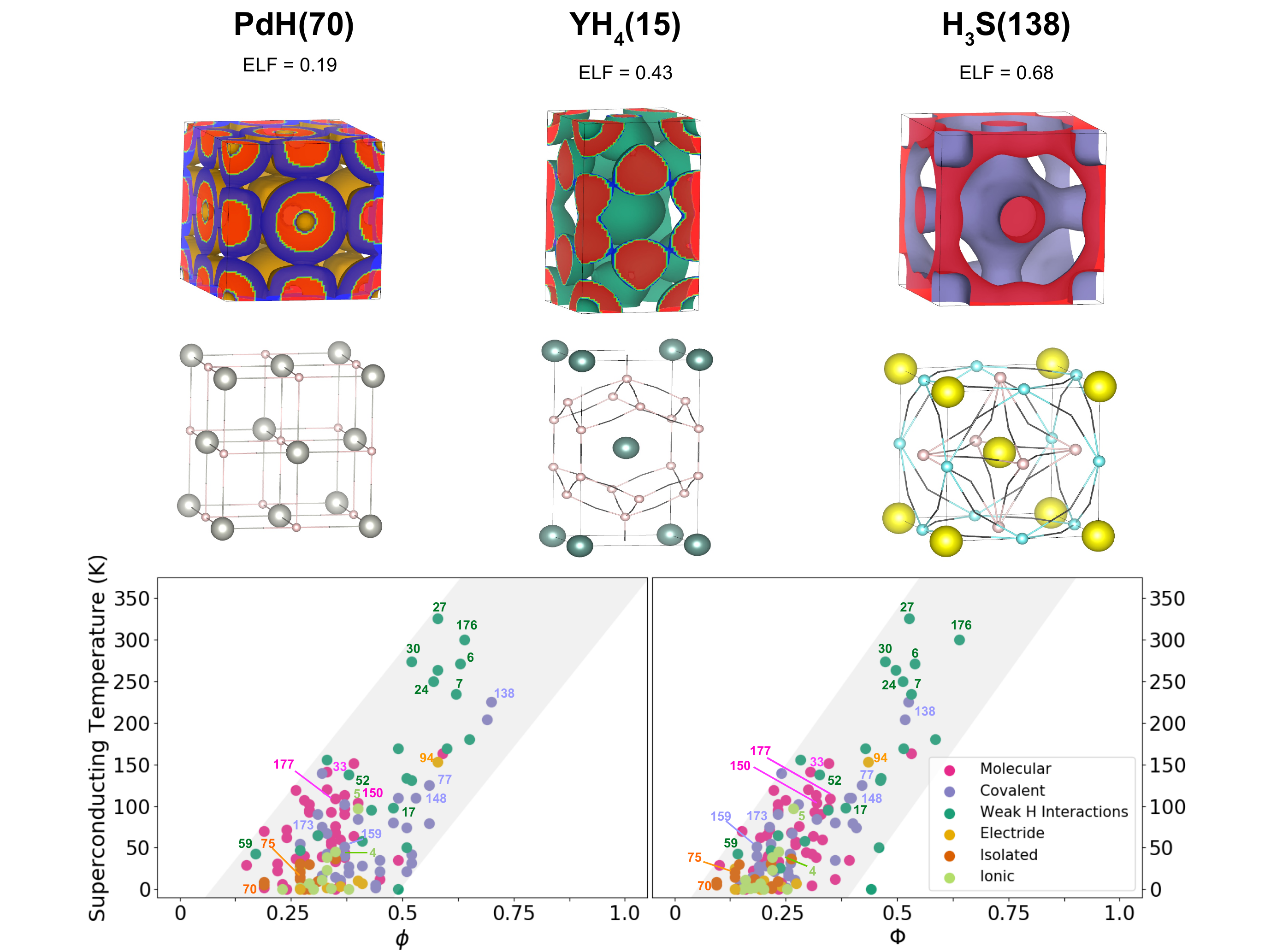}
\caption{\textbf{Networking value.} The upper panel shows the ELF isosurface and the three dimensional network spanning through all the crystal, which is formed by the ELF saddle points and the atoms, associated with the $\phi$ value for PdH(70), YH$_4$(15), and H$_3$S(138). The {\it networking value} is given for each case. The bottom panels show the critical temperature $T_c$ as a function of the networking value $\phi$ (left) and the networking value multiplied by $H_f$, $\Phi = \phi H_f$ (right). In both cases there is a linear correlation with the $T_c$, although $\Phi$ correlates slightly better.}
\label{fig:Network}
\end{figure*}

In Fig. \ref{fig:Network} we provide the ELF isosurface related to the $\phi$ value as well as the network created by the saddle points for this ELF value for PdH(70), YH$_4$(15), and H$_3$S(138). These three cases are related to the most common types of networks identified during the analysis: {\it isolated}, \textit{weak covalent hydrogen-hydrogen interaction}, and {\it covalent} families, respectively. In PdH(70) the network includes both Pd and H atoms. The isolated behavior of the atoms leads to the ELF bubbles around the atoms getting in contact at very low values of ELF, 0.19. For YH$_4$(15) the highest 3D connecting network appears at ELF=0.43, and is constructed only by hydrogen atoms showing a weak covalent interaction. This is one of the cases in which the 3D network subsisting at the highest value of ELF does not include all the atoms in the unit cell. For the case of H$_3$S(138), two interlaced networks appear at ELF=0.68 due to its high symmetry, which is supported by the weak $H=S$ covalent bonds. Another type of network is the one arising from the {\it electride} systems, where the connection appears through the isolated pockets of charge in the empty zones of the unit cell. 

Interestingly, the {\it networking value} correlates rather well with $T_c$ as shown in Fig. \ref{fig:Network}, clearly much better than any other descriptor based on the structure or the electronic properties studied so far in the literature and in Figs. \ref{fig:Group} and \ref{fig:Electronic}. We attribute this positive correlation to the capacity of this descriptor to somehow measure how the lattice vibrations affect the electronic cloud. This is not surprising, because if atoms are connected among them with highly localized electrons, phonon vibrations are prone to affect more electrons. The fact that the network is required to span through all the crystal seems also reasonable, as all, or at least many, phonon modes are expected to affect the electronic cloud.

Few attempts were made to improve the correlation between $\phi$ and $T_c$ by introducing additional conditions for the network identification, such as having all the hydrogen atoms be part of the network, connecting the atoms through short direct paths, and creating connections between close hydrogen atoms up to a given distance. The introduction of these restrictions however produced only minor improvements at the prize of complicating the definition of $\phi$. On the opposite side, relaxing the condition about the 3D nature of the network to simply two dimensional sheets or one dimensional worms worsened the correlation. This suggests that forming a 3D bonding network is crucial, which we believe it is related to the fact that in this case the networking value better captures the overall coupling of the phonon modes with the electrons as argued above.

As seen in Fig. \ref{fig:Network}, an improvement in the correlation is obtained by multiplying $\phi$ by the hydrogen fraction $H_f$:
\begin{equation}
    \Phi = \phi  H_f.
\end{equation}
A reasonable explanation for such improvement is that the hydrogen fraction is a rough estimation of the multiplicity of hydrogen bonds. Systems with few H atoms will tend to form less bonds in which H atoms participate. This is in contrast with hydrogen rich systems with an incredible number of bonds in which H atoms participate. An example is HSe(148) with just 7 bonds per formula unit and $H_f = 0.5$, in contrast with LaH$_{10}$(30) with 20 bonds per formula unit and $H_f$ = 0.90.

\begin{figure}[ht]
\includegraphics[width=\columnwidth]{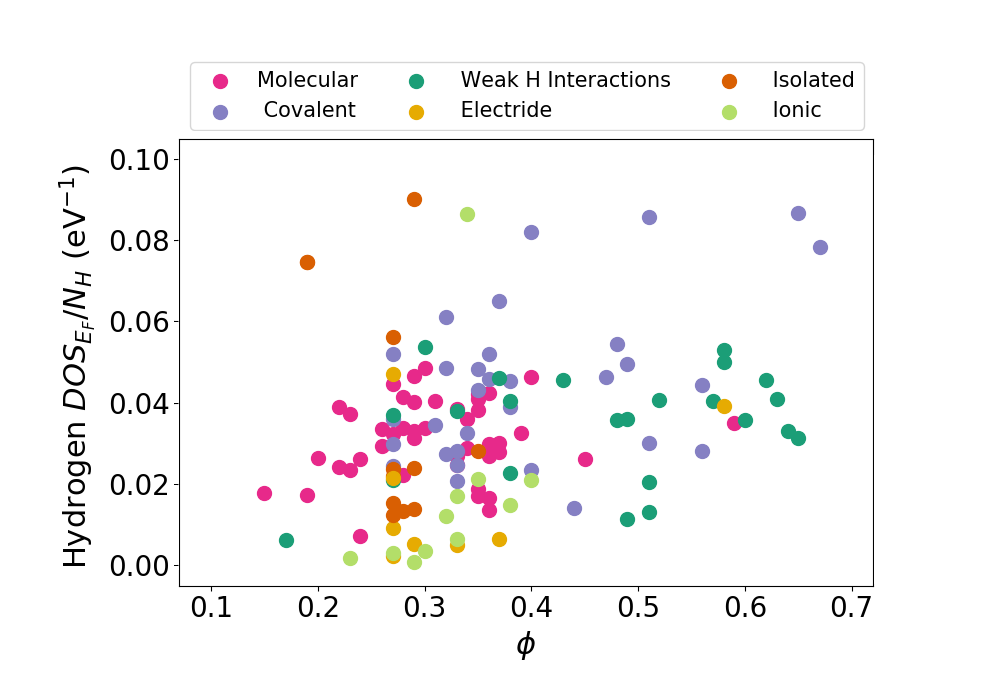}
\caption{\textbf{Networking value with respect to the DOS.} The figure shows the {\it networking value} $\phi$ as a function of the hydrogen contribution to the DOS at the Fermi energy per hydrogen atom.}
\label{fig:NetworkvsDOS}
\end{figure}

We have also investigated the possible correlation between the {\it networking value} and the DOS which was found missing (see Fig. \ref{fig:NetworkvsDOS}). The fact the {\it networking value} does not correlate at all with the DOS at the Fermi level , not even with the DOS coming from H atoms, underlines that these two descriptors are measuring different things. While $\phi$ is able to capture the effect of the electron-phonon interaction, the DOS cannot, and consequently does not correlate well with $T_c$. On the contrary, $\phi$, or its product with the hydrogen fraction $\Phi$, determine for the first time a magnitude that shows a striking correlation with $T_c$, and constitutes a first positive and cheap identification of hydrogen-based superconductors. 

\noindent

\section{Discussion}


After the large analysis presented in this work, we can conclude that the highest critical temperatures are achieved in the {\it molecular}, {\it covalent} and \textit{weak covalent hydrogen-hydrogen interaction} families. These three families are different expressions of covalent bonds, where electrons are strongly localized. The highest critical temperatures appear between groups 1 and 5 of the periodic table, where bonds are mainly driven by covalent hydrogen-hydrogen interactions, and 13 and 16, where  covalent bonds are predominantly between hydrogen and host atoms. It seems that these {\it covalent} compounds reach their highest $T_c$ values for systems with large symmetry and hydrogen-host bonds at a distance of about 1.55 \AA, without direct hydrogen-hydrogen bonding.

It is important to remark that our work shows how the {\it molecular} family transitions towards the \textit{weak covalent hydrogen-hydrogen interaction} one, with an associated increase in $T_c$. The transition is smooth, starting from low $T_c$ systems with only hydrogen molecules, such as  H$_4$I(164), going then through a mixed phase where molecules expand and intermolecular H-H interactions start to be present as for ScH$_9$(31) and ScH$_7$(34), to finally transition towards a full weak interacting behavior with no molecules as found for YH$_{10}$(27) and LaH$_{10}$(30), which have the largest {\it networking value} and $T_c$. This suggests that stretching hydrogen molecules is beneficial for superconductivity. The same conclusion is reached looking at the two pure hydrogen phases studied here (systems 176 and 177). The hydrogen phase ascribed to the \textit{weak covalent hydrogen-hydrogen interaction} behavior (176) shows a far higher $T_c$ of around 300 K compared to the purely molecular phase (177), with $T_c=109$ K. Therefore, stretching hydrogen molecules by chemical or mechanical means in systems containing many H$_2$ units seems a very promising path to discovering new high-$T_c$ compounds. This seems to put in context the extraordinary prediction of a critical temperature of 473 K in Li$_2$MgH$_{16}$ \cite{11}, where  doping a molecular MgH$_{16}$ compound with Li brakes the molecular units, transforming the system into one with \textit{weak covalent hydrogen-hydrogen interactions}.

The {\it networking value} $\phi$ defined here is able to capture effectively how sensitive the electronic cloud is on average to lattice vibrations, and, consequently, correlates well with $T_c$. The {\it networking value} improves all other structural or electronic descriptors previously studied \cite{13,15,17}. As extracting $\phi$ simply requires the analysis of ELF isosurfaces, which can be easily obtained post-processing DFT ground state calculations, it offers a simple way of screening hydrogen-based superconductors, as well as showing the correct directions to chemically engineering better hydrogen-based superconductors. Interestingly, as the definition of $\phi$ is completely general, not limited to the presence of hydrogen in the system, it could potentially be used to estimate the $T_c$ of all phonon-mediated superconductors. Eventually, it may be also worth studying it in unconventional superconductors, as the $\phi$ is only based on the analysis of the electronic cloud and correlations could also appear. 

As a final word, we would like to underline that the superconducting critical temperatures used to find correlations are extracted directly from the literature, without being recalculated (see Supplementary Data Table I to check the reference from which the $T_c$ value was taken for each case). All the $T_c$ values were obtained by first principles DFT calculations, but at different levels of theory, for instance, for the estimation of the critical temperature. The wide grey area in Fig. \ref{fig:Network} could be the result of such inconsistencies. In addition, most of these $T_c$ values have been obtained assuming that the ground state structure is the one given by the minimum of the Born-Oppenheimer energy surface (classical approximation) and that lattice vibrations can be described within the harmonic approximation around these positions. However, in hydrogen-based superconductors, recent calculations have shown that the crystal structure can be largely modified by ionic quantum effects and that anharmonicity strongly renormalizes the obtained harmonic phonon spectra, which can strongly impact the predicted $T_c$\cite{PhysRevLett.111.177002,PhysRevB.89.064302,33,34,Errea81,6,8,Bianco2018,hou2021strong}. These effects could produce a narrower correlation between $T_c$ and $\phi$ by introducing variations due to corrections on the structure and $T_c$.

\section{Methods}

\noindent
\textbf{DFT Calculations.}
We carry out our analysis on a sample of 178 compounds containing hydrogen that had been previously predicted to be superconductors in the literature. Most of the chosen compounds were those summarized in Ref. \cite{7}, which had been analyzed before in the literature \cite{scList,100,101,102,103,104,105,106,107,108,109,110,111,112,113,115,116,117,118,119,120,121,122,123,124,125,126,127,128,129,130,131,132,133,134,135,136,137,138,139,140,141,142,143,144,145,146,147,148,149,150,151,152,153,154,155,156,157,158,159,160,161,162,163,164,165,166,167,168,169,170,171,172,173,174,175,176,177,178,179,180,181,182,183,184,185,186,187,188,189,190}. We took each of these compounds, relax them classically at a given pressure with DFT at the Born-Oppenheimer minimum position, and calculated electronic properties for them. Due to the immense work required, we did not perform the $T_c$ calculations, but took the values predicted in the literature. Among the 178 compounds, 43 were discarded due to lack of information on the atomic structure, which made impossible their analysis. Supplementary Data Tables I-III summarize all the calculated data presented here, as well as from which reference the  $T_c$ value was taken.

All DFT calculations were performed with the plane-wave {\sc Quantum ESPRESSO} (QE) package \cite{27,28}. The exchange correlation potential was approximated with the Perdew-Burke-Ernzerhof parametrization \cite{GGA-PBE}. At least the first few upper core orbitals were included in the pseudopotential for the host atom. The cutoff for the wave-functions and the density were respectively 70 Ry  and 700 Ry. Integrations over the Brillouin zone were performed with the Methfessel-Paxton smearing technique \cite{PhysRevB.40.3616}, with a 0.02 Ry broadening. These integrations were performed with dense $\mathbf{k}$ point grids, where a volume of 0.001 {\AA}$^{-3}$ was occupied per $\mathbf{k}$ point in the Brillouin zone for the self-consistent calculation and a volume of 0.0002 {\AA}$^{-3}$ for the non self-consistent calculation. The electronic properties such as the ELF, the DOS, and the charge distribution were calculated for each system using the QE post-processing tools through the results obtained for the non self-consistent calculations.

\noindent
\textbf{Bader charges. }
There have been numerous approaches in order to determine the charge associated to an atom in a molecule. Probably, one of the most useful in solid state is that derived from the electron density, introduced by Bader and coworkers in what is called the Quantum Theory of Atoms in Molecules \cite{BA90}.
In an ordinary solid, the electron density has its maxima (cusps) at the nuclei and decays exponentially as the electron density moves away from the nuclei. The resulting topology looks like an assemblage of mountains, each of which is identified as an atom. 
The zero gradient surface around these maxima are well defined surfaces that lead to atoms as non-overlapping units. This allows determining their charge by mere integration of the electron density within their associated region of space. Since these regions are non-overlapping these charges have the advantageous properties of being additive.

\noindent
\textbf{The Electron Localization Function. }
The electron localization function (ELF) was developed by Becke and Edgecombe in 1990 \cite{BE90} for the analysis in real space of electron localization, and later on reinterpreted by Savin \cite{KS96} in terms of the Pauli kinetic energy density ($t_p$) corrected by the homogeneous electron gas kinetic energy density ($t_w$):
\begin{eqnarray}
\chi=\frac{1}{1+(t_p/t_w)^2}.
\label{eta}
\end{eqnarray}
$\chi$ is then mapped to run from 0 to a maximum value of 1:
\begin{eqnarray}
ELF=\frac{1}{1+\chi^2}.
\label{elf}
\end{eqnarray}
Values close to 1 appear in those places where electrons are localized. Hence, maxima appear in the bonds (as well as in cores and lone pairs).  
It should be noted that hydrogen constitutes a peculiar case. A molecule such as N$_2$ will feature a maximum in the middle of the inter-nitrogen distance associated with the N-N bond and separated from the N cores. Since a hydrogen molecule only has 2 electrons, a maximum does not appear for the H-H bond, but rather a surface with very high ELF that encapsulates the H$_2$ molecule. 

It is easy to see from Eq. \ref{elf} that the value $ELF=$0.5 is associated with the distribution in a homogeneous electron gas of the same density as the point of study. Indeed, metals are characterized by very flat ELF profiles slightly deviating from 0.5.

Whereas the maxima of ELF provide a measure of how localized electrons are, its value in between these maxima characterize delocalization in between these regions, i.e. how easy it is for electrons to go from one localized unit to another \cite{deloc}. 
When the ``easiest pathway'' is analyzed in the full crystalline cell, we end up with a picture of how easy it is for electrons to move across the crystal.

\section{Data availability}
All the results presented in this work are summarized in Supplementary Data Tables I-III. Further data or details are available from the corresponding author upon reasonable request. 

\section{Code availability}
Quantum ESPRESSO is an open-source suite of computational tools available at https://www.quantum-espresso.org.

\def\bibsection{\section*{\refname}} 

\section{Acknowledgements}
This research was supported by the European Research Council (ERC) under the European Unions Horizon 2020 research and innovation programme (grant agreement No. 802533).

\section{Author contributions}
F.B. performed all the DFT calculations. I.E. conceived the project. All authors contributed to the analysis of the data and to the writing of the manuscript.

\section{Competing interests}
The authors declare no competing interests.


\clearpage

\end{document}


\title{Supplementary Information: \\ Strong correlation between bonding network and critical temperature in hydrogen-based superconductors}

 
\author{Francesco Belli}
\affiliation{\CFM} \affiliation{\Unidonostia}

\author{J. Contreras-Garcia}
\affiliation{\Sorbonne}

\author{Ion Errea}
\affiliation{\CFM} \affiliation{\Unidonostia} \affiliation{\DIPC}

\maketitle

 
\tableofcontents

\renewcommand{\tablename}{\textbf{Supplementary Data Table}}

\section{Supplementary Data Tables}

\begin{longtable}{c|c|c|c|c|c|c}
    \caption{Table reporting item number used in the manuscript, chemical formula, space group, superconducting critical temperature as predicted, pressure at which it has has been predicted, hydrogen fraction $H_f$, and the reference from which the $T_c$ value has been extracted. The empty cells refer to missing values from the literature or to the impossibility to perform calculations.} \\
    \label{table:Sup1}
Item	&	Chemical formula  	&	Space group	&	$T_c$ (K)	&	Pressure (GPa)	&	$H_f$	&	Reference	\\ 
\hline
0    &  LiH$_{2}$    &  $  P4/mbm      $ &     0           &    150        &   0.666    &    \cite{100}   \\	
1    & LiH$_{2}$    &  $  R\bar{3}m        $ &     38.34       &    140        &   0.857    &    \cite{100}  \\	
2    & LiH$_{8}$    &  $  I422        $ &     31.04       &    200        &   0.888    &    \cite{100} \\	
3    & KH$_{6}$    &  $  C2/c        $ &     69.8        &    166        &   0.8      &    \cite{101} \\	
4    & BeH$_{2}$    &  $  Cmcm        $ &     45.1        &    204        &   0.666    &   \cite{102}\\	
5    & BeH$_{2}$    &  $  P4/nmm      $ &     97          &    349        &   0.666    &   \cite{102}\\
\end{longtable}

\clearpage

\begin{longtable}{c|c|c|c|c}
\caption{Table reporting the item index, chemical formula, {\it networking value} $\phi$, dominant bonding family assigned, and shortest H-H distance for all the compounds. The empty cells refer to missing values from the literature or to the impossibility to perform calculations. } \\
\label{table:Sup2} 
Item & Chemical formula  & {\it Networking value} $\phi$ & Dominant bonding family &  H-H Distance (\AA) \\
\hline
0    & LiH$_{2}$        &   0.24   &   Molecular                &       0.759                 \\
1    & LiH$_{2}$        &   0.37   &   Molecular                &       0.803                 \\
2    & LiH$_{8}$        &   0.22   &   Molecular                &       0.844                 \\
3    & KH$_{6}$        &   0.19   &   Molecular                &       0.887                 \\
4    & BeH$_{2}$        &   0.35   &   Ionic                    &       1.404                 \\
5    & BeH$_{2}$    	 &   0.4    &   Ionic                      &         1.368                 \\
\end{longtable}                  

\clearpage

\begin{longtable}{c|c|c|c|c}
\caption{ Table reporting the item index, chemical formula, total DOS per electron, hydrogen contribution to the DOS per hydrogen atom, and hydrogen fraction of the DOS at the Fermi energy. The empty cells refer to missing values from the literature or to the impossibility to perform calculations.} \\
\label{table:Sup3} 
Item & Chemical formula & Total DOS ($eV^{-1}$) & Hydrogen DOS ($eV^{-1}$) & Hydrogen DOS  fraction \\
\hline
0    & LiH$_{2}$    &                         0.00298      &       0.00709      &       0.94924    \\
1    & LiH$_{2}$    &                         0.02220      &       0.02933      &       0.88063    \\
2    & LiH$_{8}$    &                         0.03063      &       0.039        &       0.92588    \\
3    & KH$_{6}$     &                         0.05004      &       0.1429       &       0.87854    \\
4    & BeH$_{2}$    &                         0.01675      &       0.02111      &       0.42005    \\
5    & BeH$_{2}$    &                         0.01743      &       0.02091      &       0.39974    \\
\end{longtable}


\includegraphics[width=0.1\textwidth]{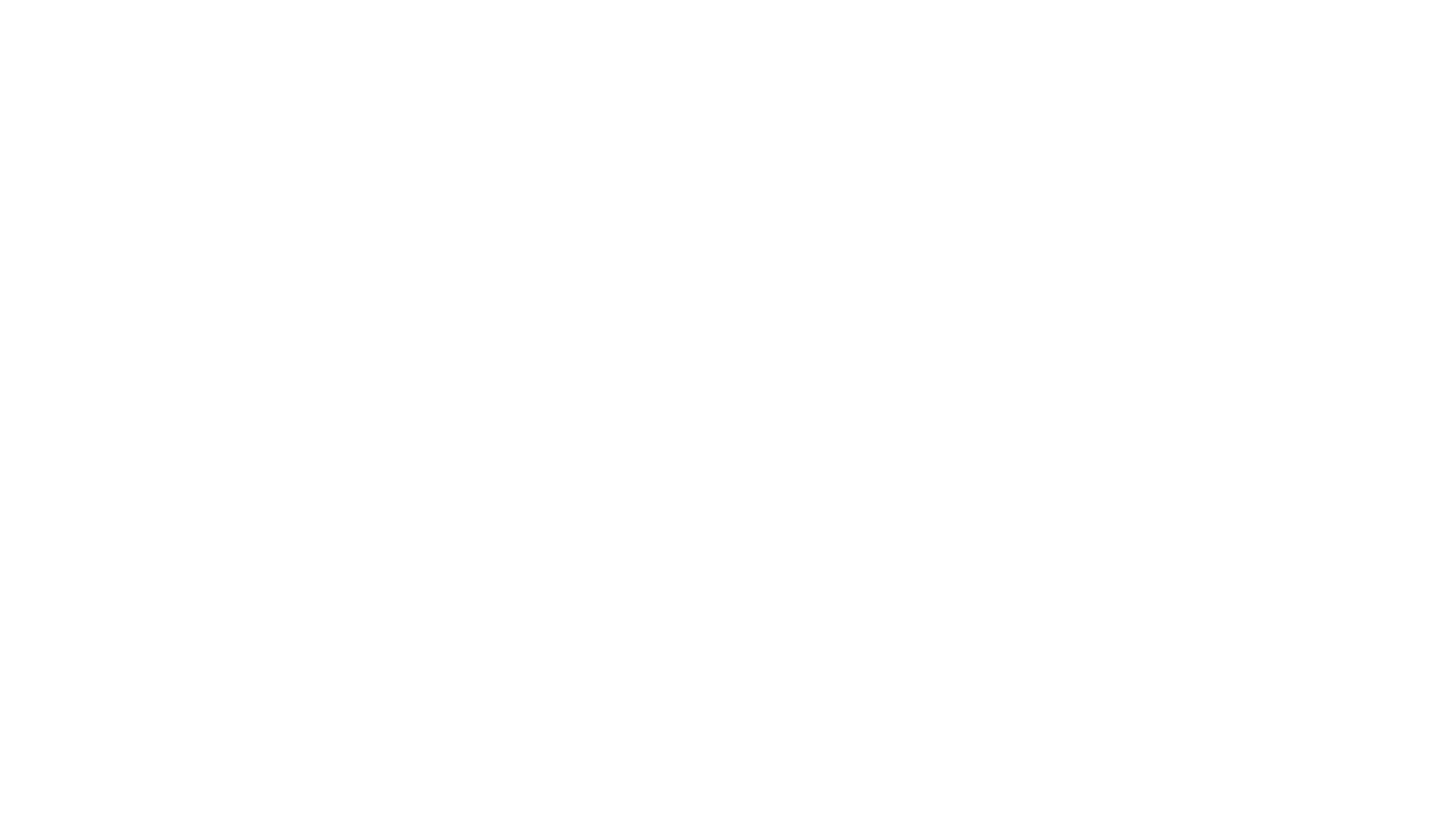}